\documentclass{article}
\usepackage{LaThuileFPSpro}
\usepackage{graphicx}

\begin{document}
\newcommand{\dzero}     {D\O\ }
\newcommand{\ttbar}     {$t\overline{t}$\ }
\newcommand{\ppbar}     {$p\overline{p}$\ }
\newcommand{\qqbar}     {$q\overline{q}$\ }
\newcommand{\MET}       {\mbox{$\not\!\!E_T$}\ }
\newcommand{\METns}     {\mbox{$\not\!\!E_T$}}
\newcommand{\pt}        {$p_T$\ }
\newcommand{\wrt}       {with respect to\ }
\renewcommand{\labelenumi}{\roman{enumi})}

\title{TOP QUARK PAIR PRODUCTION AND PROPERTIES MEASUREMENTS AT THE TEVATRON}
\author{Marc-Andr\'e Pleier\\
  {\em  University of Bonn, 53115 Bonn, Germany}\\
  On behalf of the CDF and \dzero Collaborations}
\maketitle

\baselineskip=11.6pt

\begin{abstract}
  The Tevatron proton-antiproton collider at Fermilab with its centre
  of mass energy of 1.96~TeV is currently the only source for the
  production of top quarks. This report reflects the current status of
  measurements of the top quark pair production cross section and
  properties performed by the CDF and \dzero Collaborations. Utilising
  datasets of up to two fb$^{-1}$, these measurements allow
  unprecedented precision in probing the validity of the Standard
  Model.
\end{abstract}
\newpage
\baselineskip=14pt

\section{Introduction}
\label{sec:Intro}
Since its discovery in 1995 at the Tevatron\cite{topdiscovery}, the
top quark remains the heaviest known fundamental particle to date.
With a mass of 172.6 $\pm$ 1.4~GeV/c$^2$\cite{topmass}, it is
considered to be intimately connected with the mechanism of
electroweak symmetry breaking in the Standard Model of elementary
particle physics (SM) and also to be sensitive to physics beyond the
framework of the SM.

This article reports recent measurements by the CDF and \dzero
Collaborations that probe the SM expectations for deviations both in
the production and decay of the top quark. After a brief outline of
the top quark properties within the SM framework in
Section~\ref{sec:SMtop}, the current status of measured top quark pair
production rates is given Section~\ref{sec:xsec}, followed by a
section on searches for top quark production beyond the SM. The
subsequent three sections describe measurements probing the top quark
decay in terms of branching fractions, search for flavour-changing
neutral currents and the helicity of the $W$ boson in the top quark
decay, respectively. A conclusion is given in the final
Section~\ref{sec:conclusion}.

\section{Top Quark Pair Production and Decay in the SM}
\label{sec:SMtop}
Within the framework of the SM, top quark production at the Tevatron
proceeds mainly in pairs: $p\bar{p}\to t\bar{t} + X$ via the
strong interaction (85\% \qqbar annihilation and 15\% gluon-gluon
fusion).

The corresponding production cross section has been evaluated at
next-to-leading order (NLO) QCD using two different approaches: One
calculation considers soft gluon corrections up to
next-to-next-to-next-to leading logarithmic (NNNLL) terms and some
virtual terms in a truncated resummation, yielding 6.77 $\pm$ 0.42 pb
for a top quark mass of 175 GeV/c$^2$\cite{ttbar_xsecKid}, while
another calculation using the NLO calculation with LL and NLL
resummation at all orders of perturbation theory gives
6.70$^{+0.71}_{-0.88}$ pb for a top quark mass of 175
GeV/c$^2$\cite{ttbar_xsecCac}. If a PDF uncertainty is combined
linearly with the theoretical uncertainty for the first result --
similar to what is done for the second result -- both predictions
exhibit not only similar central values but also similar relative
uncertainties of $\approx$12-13\%.

Due to its large mass, the top quark has an extremely short lifetime
of approximately \mbox{$5\cdot 10^{-25}$~s}, which makes it decay
before it can form hadrons -- a unique feature setting it apart from
all other quarks. Since the top quark mass is well above the threshold
for $Wq$ decays with $q$ being one of the down-type quarks $d, s, b$,
this two-body decay dominates the top quark decay. As each quark
flavour contributes to the total decay rate proportional to the square
of the respective CKM matrix element $V_{tq}$, top decays into $Ws$
and $Wd$ are strongly suppressed \wrt the dominant decay $t\to Wb$.

Consequently, top quark pair events contain a $b$ and a $\bar{b}$
quark from the \ttbar decay, and depending on the decay modes of the
two $W$ bosons, the observed top quark pair final states can be
divided into three event classes:
\begin{enumerate}
\item In {\em dilepton} events, both $W$ bosons decay leptonically,
resulting in a final state containing two isolated high-\pt leptons,
missing transverse energy \MET corresponding to the two neutrinos and
two jets. This final state constitutes $\approx$5\% of the \ttbar
events (not counting $\tau$ leptons) and gives the cleanest signal but
suffers from low statistics.

\item In {\em lepton+jets} events, one $W$ boson decays leptonically,
the other one hadronically, resulting in one isolated high-\pt lepton,
\MET and four jets. Events in the $e$+jets or $\mu$+jets channels
yield $\approx$29\% of the branching fraction ($\approx$34\% when
including leptonic $\tau$ decays) and provide the best compromise
between sample purity and statistics.

\item In {\em all-hadronic} events, both $W$ bosons decay to
$\overline{q}q'$ pairs, resulting in a six-jet final state. With a branching fraction
of $\approx$46\%, this final state represents the biggest fraction of
\ttbar events, but it is also difficult to separate from the large
background of multijet production.

\end {enumerate}
All of these final states contain two $b$-jets from the hadronisation
of the (anti-) $b$ quarks, and additional jets can arise from initial
and final state radiation.

\section{Measurement of the Top Quark Pair Production Cross Section}
\label{sec:xsec}
Top quark pair production cross section measurements provide a unique
test of the predictions from perturbative QCD calculations at high
transverse momenta. Analysing all three event classes allows both the
improvement of statistics of top events and studies of properties and
important checks for physics beyond the SM that might result in
enhancement/depletion in some particular channel via novel production
mechanisms or decay modes.

The following subsections give an overview of the cross section
measurements pursued at the Tevatron rather than quoting single cross
section results, with the exception of the most precise single
measurement to date, obtained by \dzero in the lepton+jets channel.
All current measurements are summarised in Figure~\ref{fig:xsecsummary}.

\subsection{Dilepton Final State}
A typical selection of dilepton events requires two isolated high \pt
leptons, \MET and at least two central energetic jets in an event. The
most important physics background processes containing both real
leptons and \MET are Z/$\gamma^{*}$+jets production with
$Z/\gamma^{*}\to\tau^{+}\tau^{-}, \tau\to e,\mu$ and the production of
dibosons ($WW, ZZ, WZ$). Instrumental backgrounds are to be considered
as well, arising from misreconstructed \MET due to resolution effects
in Z/$\gamma^{*}$+jets production with $Z/\gamma^{*}\to
e^{+}e^{-}/\mu^{+}\mu^{-}$, and also from $W$+jets and QCD multijet
production where one or more jets fake the isolated lepton signature.
To ensure proper description of the instrumental backgrounds, these
are usually modelled using collider data, while for the physics
backgrounds typically Monte Carlo simulation is used.

A further enhancement of the signal fraction in the selected data
samples is possible by requiring additional kinematical event
properties like the scalar sum of the jet $p_T$s $H_{T}$ to be above a
certain threshold or rejecting events where both selected leptons have
like-sign electric charge. The obtained purities in such selected
samples are usually quite good with a signal to background ratio (S/B)
better than 2 at least, although signal statistics are low. The
acceptance for dilepton final states can be enhanced by loosening the
selection to require only one fully reconstructed isolated lepton
($e,\mu$) in addition to an isolated track (``$\ell$+track
analysis''). In particular, such a selection allows the inclusion of
``1~prong'' hadronic $\tau$ decays.

The top quark production cross section was recently measured for the
first time also in the lepton+tau final state by D\O\cite{leptauxsec},
using events with hadronically decaying isolated taus and one isolated
high \pt electron or muon. To separate real taus from jets, a neural
network was used, and the sample purity was enhanced by requiring
$b$-jet identification (see Section~\ref{lpjxsec}) in the selected events. The result is shown
together with the other measurements in Figure~\ref{fig:xsecsummary}.

\subsection{Lepton+Jets Final State}
\label{lpjxsec}
\begin{figure}[!t]
  \begin{center} 
    \includegraphics[width=.48\textwidth]{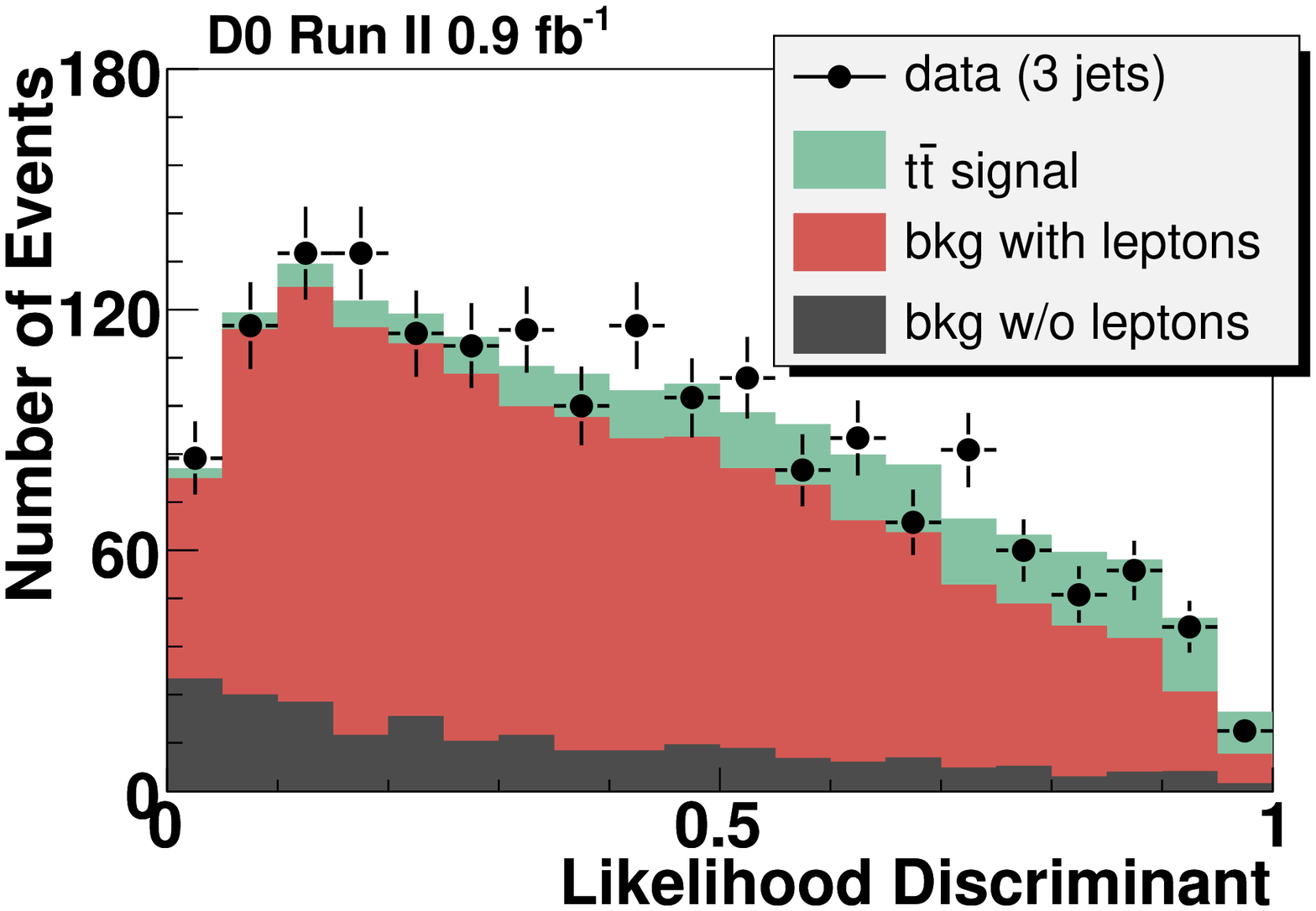}\hspace*{1mm}
    \includegraphics[width=.48\textwidth]{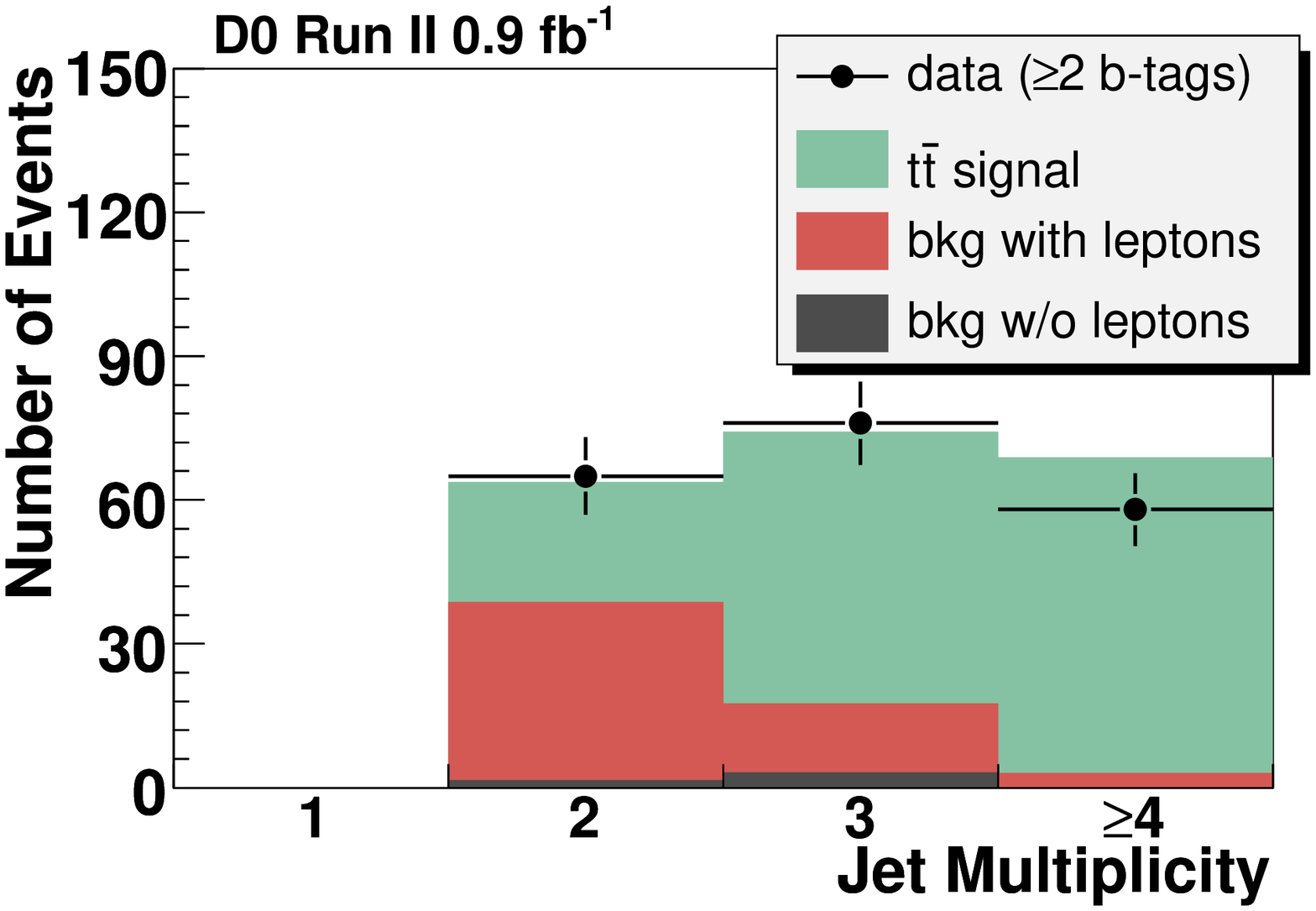} 
    \caption{\it Sample composition in a lepton+jets sample, requiring
      three jets (left) or at least two $b$-tagged jets
      (right)\cite{ljetsxsec}.}
      \label{fig:ljetsxsec}
  \end{center}
\end{figure}
A typical lepton+jets selection requires one isolated high \pt lepton
($e$ or $\mu$ which includes $\tau\to e\nu\bar{\nu}$, $\tau\to\mu\nu\bar{\nu}$), \MET and at least 4 jets, yielding samples with a S/B
around 1/2. The dominant physics background to be considered here
comes from $W$+jets production while the main instrumental background
arises from QCD multijet production where a jet fakes the isolated
lepton signature.

The cross section can be extracted from such a selected sample either
purely based on topological and kinematical event properties combined
in a multivariate discriminant to separate the \ttbar signal from
background or by adding identification of $b$-jets. Since topological
analyses do not depend on the assumption of 100\% branching of $t
\rightarrow Wb$, they are less model-dependent than tagging analyses.
On the other hand, requiring $b$-jet identification is a very powerful
tool in suppressing the background processes, which typically exhibit
little heavy flavour content. With $b$-jet identification, the top
signal can also be easily extracted from lower jet multiplicities,
where topological analyses need to impose additional selection
criteria like cutting on $H_T$ to be able to extract the signal. In
addition, $b$-tagged analyses can provide very pure signal samples,
easily exceeding a S/B $>$ 10 in selections requiring at least four
jets with two identified b-jets (see for example Figure
\ref{fig:ljetsxsec}).

The identification of $b$-jets can be based on the long {\it lifetime}
of B hadrons resulting in significantly displaced secondary vertices
with respect to the primary event vertex or large significant impact
parameters of the corresponding tracks. A combination of this type of
information in a neural network tagging algorithm yields $b$-tagging
efficiencies of about 54\% while only about 1\% of light quark jets
are misidentified as $b$-jets -- hence the improved S/B in tagged
analyses. Another way to identify $b$-jets is to reconstruct {\it soft
leptons} inside a jet originating from semileptonic B decays. So far
only soft-$\mu$ tagging has been deployed in \ttbar analyses.

The most precise \ttbar production cross section measurement to date
with a relative uncertainty of 11\% has been performed by \dzero on
0.9~fb$^{-1}$ of data in the lepton+jets channel\cite{ljetsxsec}. For
this measurement, two complementary analyses based on a kinematic likelihood discriminant and
on $b$-tagging (see Figure~\ref{fig:ljetsxsec})
were combined and yield $\sigma_{t\overline{t}} =
7.42\pm0.53\mbox{(stat)}\pm0.46\mbox{(syst)}\pm 0.45
\mbox{(lumi)}\mbox{ pb}$ for a top quark mass of 175 GeV/c$^2$.
Comparing this measurement with the theory prediction, the top quark
mass can be extracted as well, yielding $170 \pm 7$~GeV/c$^{2}$ in good
agreement with the world average.

A first $\tau$+jets cross section analysis using events with
hadronically decaying isolated taus and lifetime $b$-tagging was
performed as well by \dzero -- the result is shown together with other
measurements in Figure~\ref{fig:xsecsummary}.

\subsection{All-Hadronic Final State}
The all-hadronic final state is studied by requiring events with at
least six central energetic jets and no isolated high \pt leptons. Due
to the overwhelming background from QCD multijet production with a
cross section orders of magnitude above that of the signal process,
$b$-jet identification is mandatory for this final state. Further
separation of signal and background is achieved by using multivariate
discriminants based on topological and kinematical event properties.

\subsection{Summary of the Top Quark Pair Production Cross Section Measurements}
\begin{figure}[!t]
  \begin{center} 
    \includegraphics[width=.48\textwidth, height=55mm]{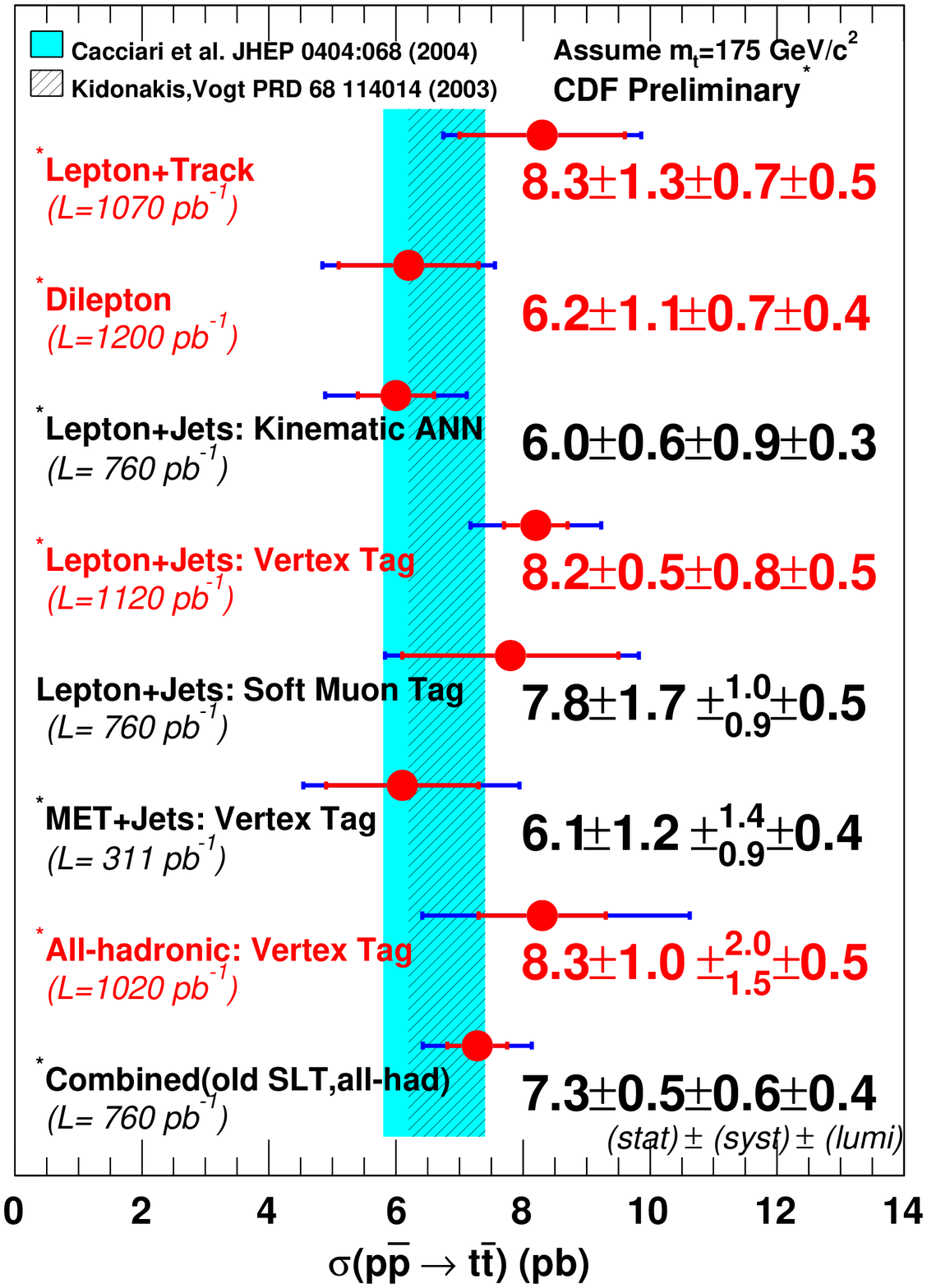} 
    \includegraphics[width=.48\textwidth, height=55mm]{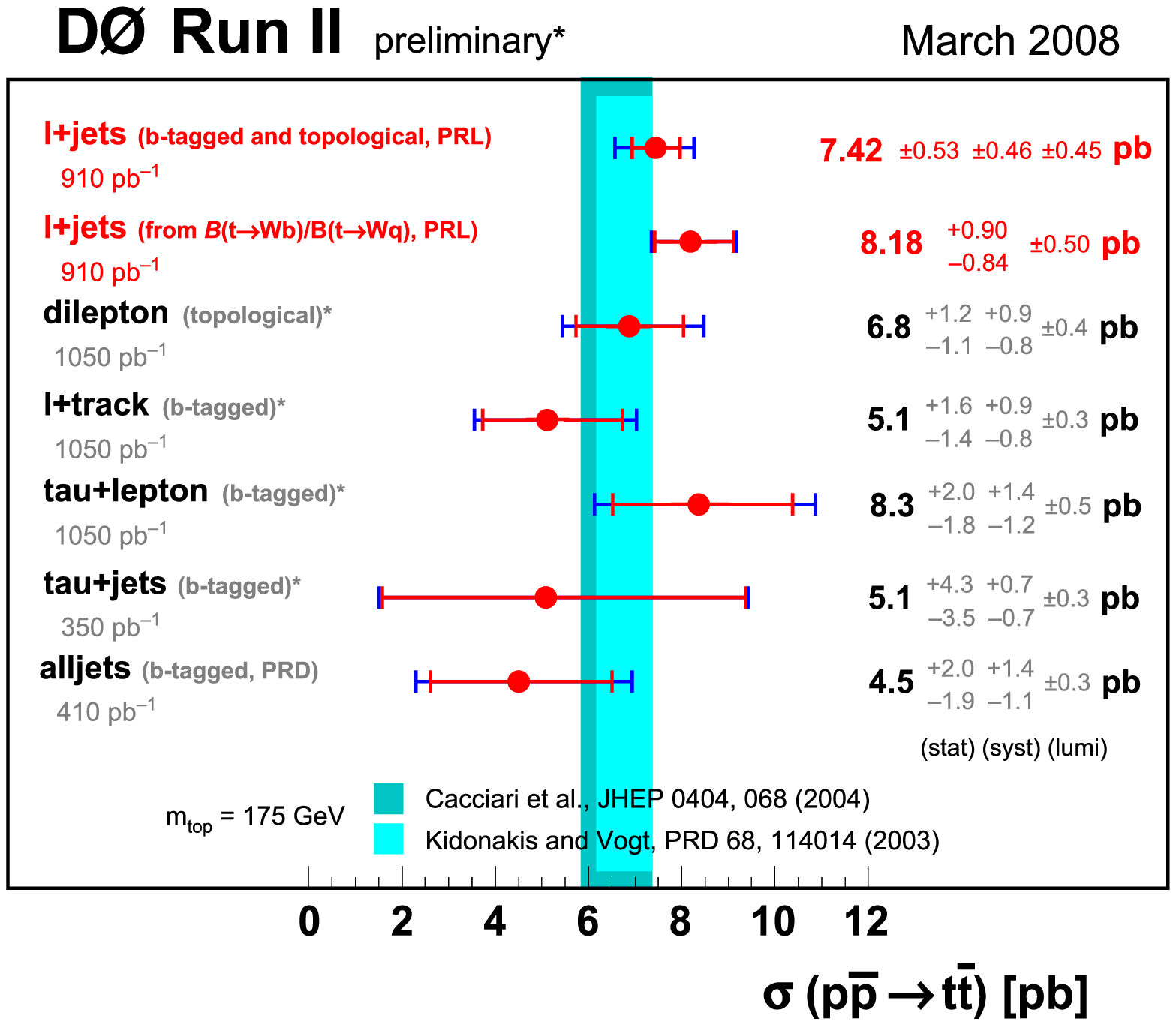}\hspace*{1mm}
    \caption{\it Top quark pair production cross section measurements
    performed by CDF and D\O.}
      \label{fig:xsecsummary}
  \end{center}
\end{figure}
Figure~\ref{fig:xsecsummary} provides an overview of recent cross
section measurements performed by CDF and D\O. All measurements show
good agreement with the SM prediction and with each other. The single
best measurements are approaching a relative precision of
$\Delta\sigma / \sigma$~=~10\% that should be achievable for the
datasets of 2~$fb^{-1}$ already at hand and provide stringent tests to
theory predictions.With increasing datasets, these measurements
naturally start to become limited by systematic uncertainties rather
than statistical ones, but it will be possible to further constrain
the systematic uncertainties as well using additional data.

Cross section measurements form the foundation for all further
property analyses like the ones described in the subsequent sections
of this article by characterising the datasets enriched in top quark
pairs and providing the necessary understanding of object
identification, background modelling and sample composition.

\section{Search for Top Quark Pair Production beyond the SM}
\label{sec:BSMprod}
\subsection{Search for a Narrow-Width Resonance decaying into \ttbar}
Various beyond the SM theories predict the existence of a massive
$Z$-like boson that could decay into \ttbar and hence add a resonant
production mode to the SM process. Any such additional production
would be visible in the \ttbar invariant mass distribution provided
the resonance $X$ decaying to \ttbar is sufficiently heavy and narrow.

\begin{figure}[!t]
  \begin{center} 
    \includegraphics[width=.48\textwidth]{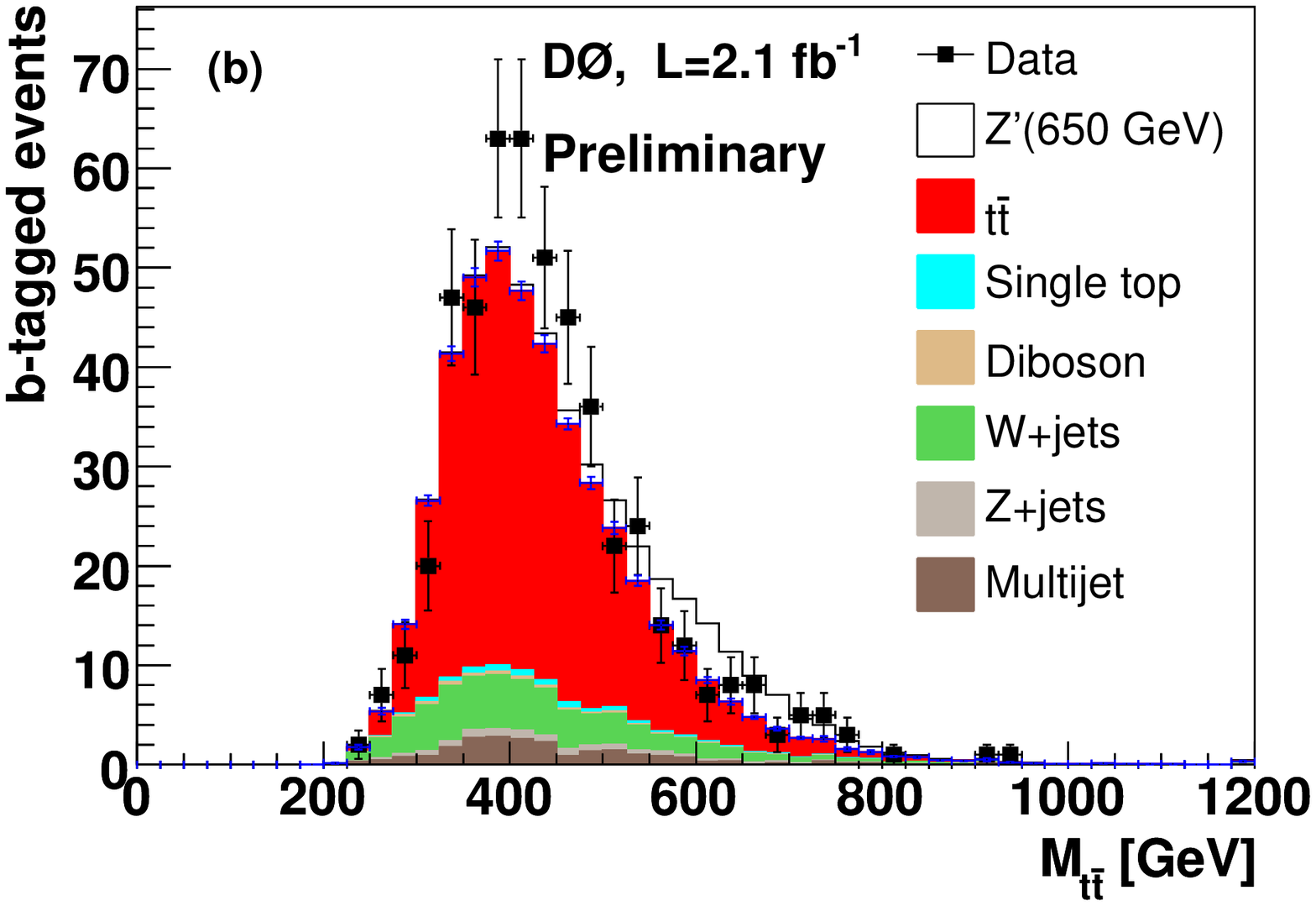}\hspace*{1mm}
    \includegraphics[width=.48\textwidth]{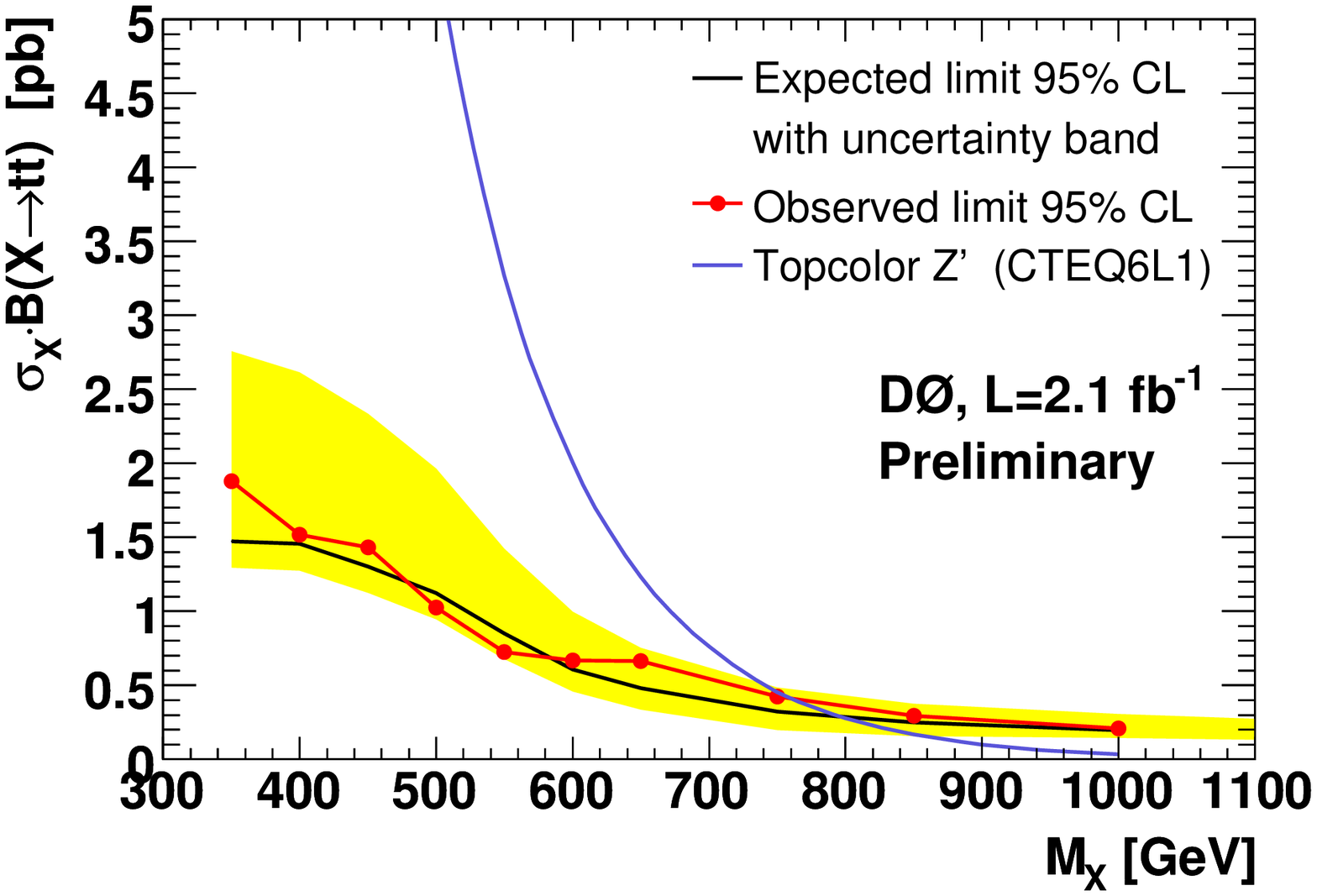} 
    \caption{\it Left: Expected and observed $ t\bar{t}$ invariant mass
    distribution in lepton+jets data with four or more jets. Right:
    Expected and observed 95\% C.L. upper limits on
    $\sigma_X\cdot{\cal B}(X\to t\bar{t})$\cite{D0Zprime}.}
      \label{fig:ttresonance}
  \end{center}
\end{figure}
Both CDF and \dzero perform a search for a generic heavy resonance $X$
of narrow width ($\Gamma_X=0.012M_X$) compared to the detector mass
resolution in $b$-tagged lepton+jets datasets. The \ttbar invariant
mass spectrum is reconstructed using either a kinematic fit to the
\ttbar production hypothesis (CDF) or directly from the four-momenta
of the up to four leading jets, the lepton and the neutrino momentum
(D\O). The latter approach was shown to provide better sensitivity for
large resonance masses than the previously used constrained kinematic
fit and also allows the inclusion of data with fewer than four jets in
case that jets merged. As both experiments observe no significant
deviation from the SM expectation, 95\% C.L. upper limits on
$\sigma_X\cdot{\cal B}(X\to t\bar{t})$ are given for values of $M_X$
between 450 and 900 GeV/c$^2$ (CDF) respectively 350 and 1000
GeV/c$^2$ (D\O, see Figure~\ref{fig:ttresonance}).

Both experiments provide 95\% C.L. mass limits for a leptophobic
top-colour-assisted technicolour $Z'$ boson as a benchmark model.
Using 955~pb$^{-1}$, CDF finds $M_{Z'}>$~720 GeV/c$^2$ (expected
limit: 710 GeV/c$^2$)\cite{CDFZprime} while \dzero finds $M_{Z'}>$~760
GeV/c$^2$ (expected limit: 795 GeV/c$^2$)\cite{D0Zprime} using 2.1
fb$^{-1}$ of data.

\subsection{Search for \ttbar Production via a Massive Gluon}
Instead of a new colour singlet particle decaying into \ttbar as
described in the previous subsection, there could also be a new
massive colour octet particle $G$ contributing to \ttbar production.
Such a ``massive gluon'' production mode would interfere with the
corresponding SM production process.

Assuming a SM top decay, CDF has performed a search for a
corresponding contribution by comparing the \ttbar invariant mass
distribution in a 1.9~fb$^{-1}$ $b$-tagged lepton+jets dataset with
the SM expectation. As the largest discrepancy with respect to the SM
observed is 1.7$\sigma$ for the explored mass and width range 400
GeV/c$^2$ $\leq M_{G} \leq$ 800 GeV/c$^2$, 0.05 $\leq \Gamma_{G}/M_{G}
\leq $ 0.5, upper and lower limits are provided on the corresponding
coupling strengths of the massive gluon\cite{massiveG}.

\subsection{Measurement of the \ttbar Differential Cross Section $d\sigma/dM_{t\bar{t}}$}
\begin{figure}[!t]
  \begin{center} 
    \includegraphics[width=.48\textwidth, height=45mm]{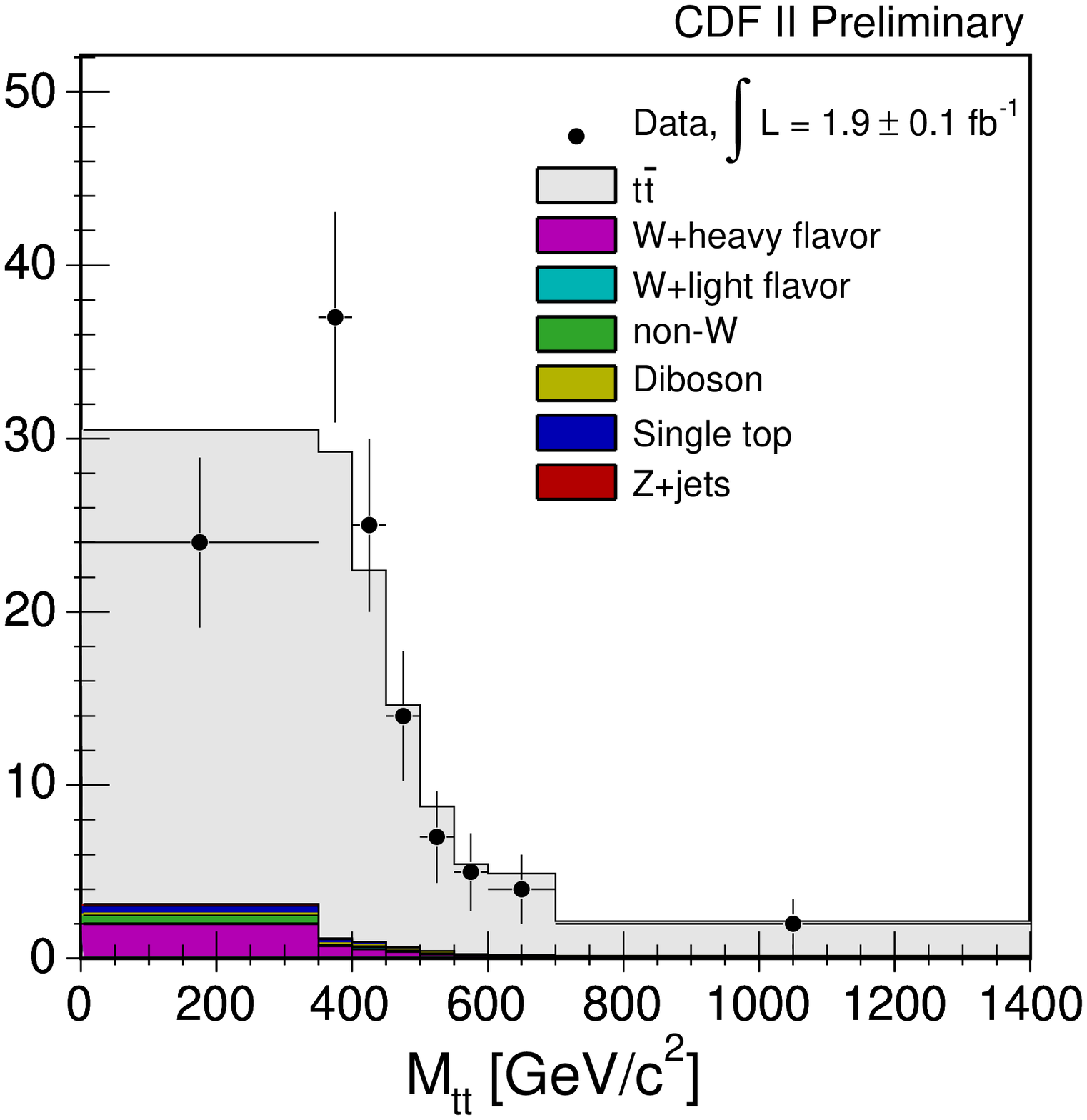}\hspace*{1mm}
    \includegraphics[width=.48\textwidth, height=45mm]{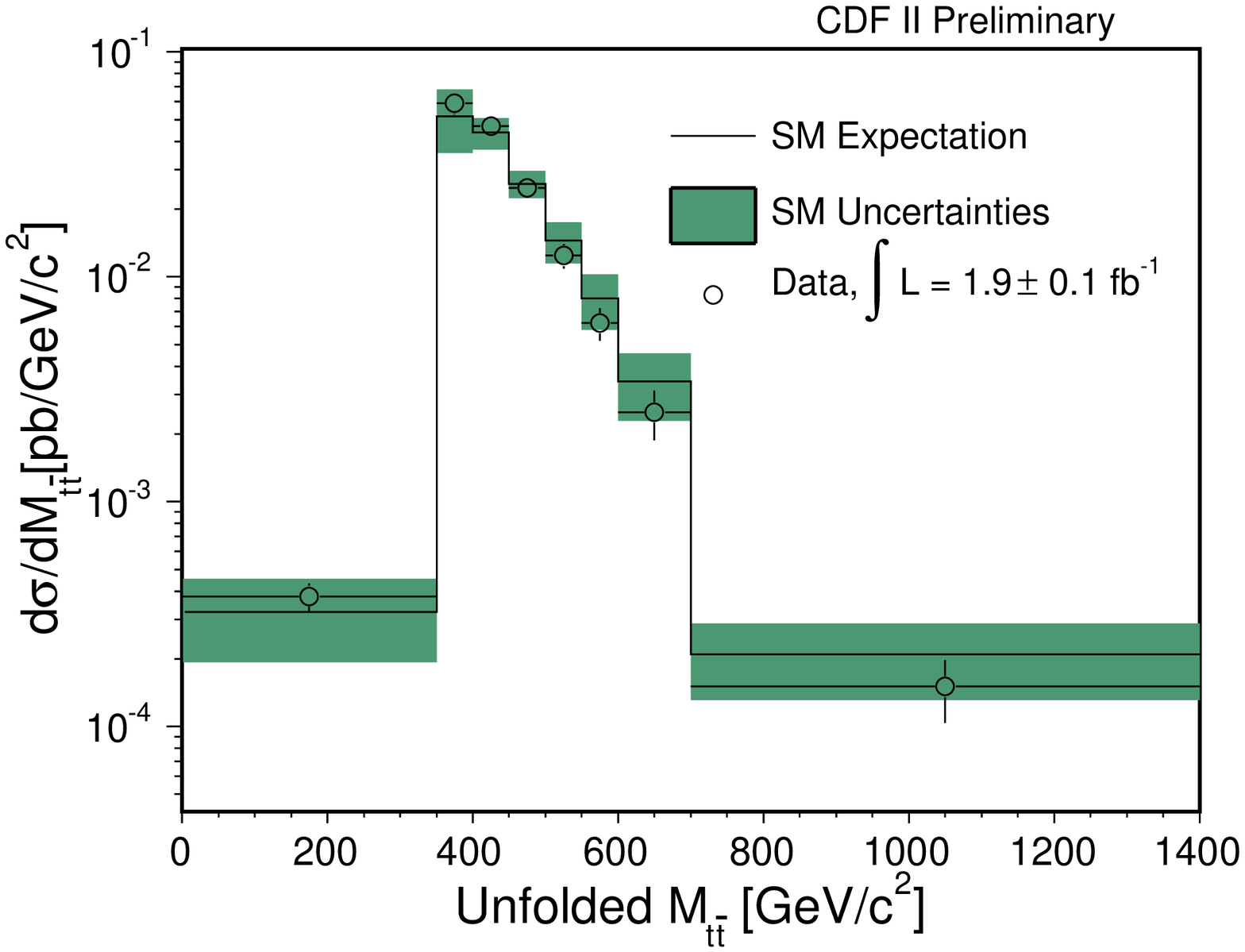} 
    \caption{\it Left: Reconstructed \ttbar invariant mass
      distribution in lepton+jets data with at least four jets. Right:
      Corresponding observed \ttbar differential cross section,
      compared to the SM expectation\cite{diffxsec}.}
      \label{fig:diffxsec}
  \end{center}
\end{figure}
Contributions beyond the SM in \ttbar production could manifest
themselves in either resonances, broad enhancements or more general
shape distortions of the \ttbar invariant mass spectrum. A very
generic way to search for such effects is to measure the \ttbar
differential cross section $d\sigma/dM_{t\bar{t}}$ and compare the
shape with the SM expectation.

CDF reconstructs the \ttbar invariant mass spectrum in a 1.9~fb$^{-1}$
$b$-tagged lepton+jets dataset (see Figure~\ref{fig:diffxsec}) by
combining the four-vectors of the four leading jets, lepton and
missing transverse energy. After subtracting the background processes,
the distortions in the reconstructed distribution due to detector
effects, object resolutions and geometric/kinematic acceptance are
corrected for by the application of a regularised unfolding technique.
From the unfolded distribution, the \ttbar differential cross section
$d\sigma/dM_{t\bar{t}}$ is extracted and its shape is compared with
the SM expectation. The shape comparison yields good agreement with
the SM, yielding an Anderson-Darling p-value of 0.45\cite{diffxsec}.

\section{Measurement of $\bf {\cal {B}}(t\rightarrow Wb )/{\cal{B}}(t\rightarrow Wq )$}
\label{sec:R}
\begin{figure}[!t]
  \begin{center} 
    \includegraphics[width=.48\textwidth]{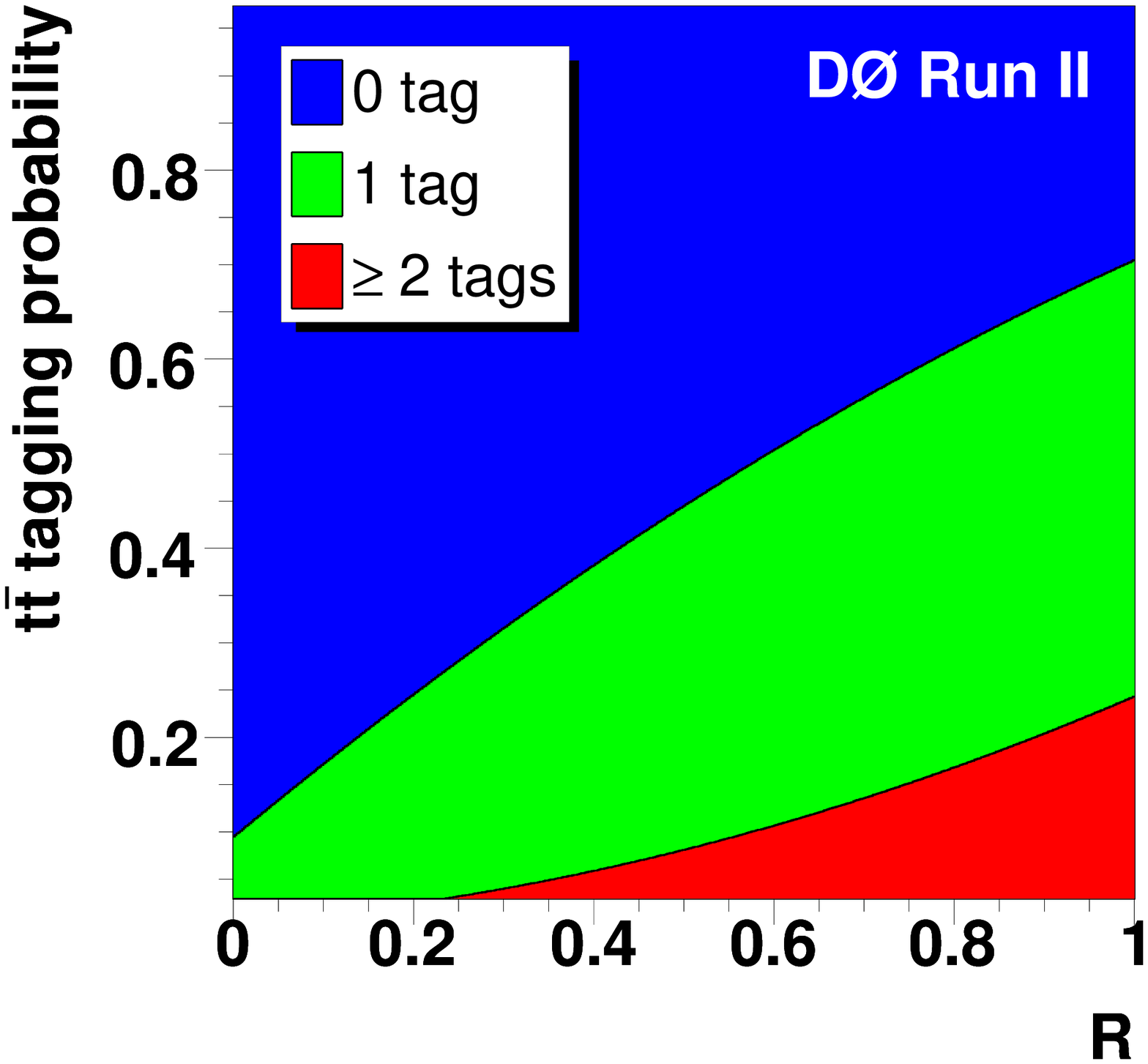}\hspace*{1mm}
    \includegraphics[width=.48\textwidth]{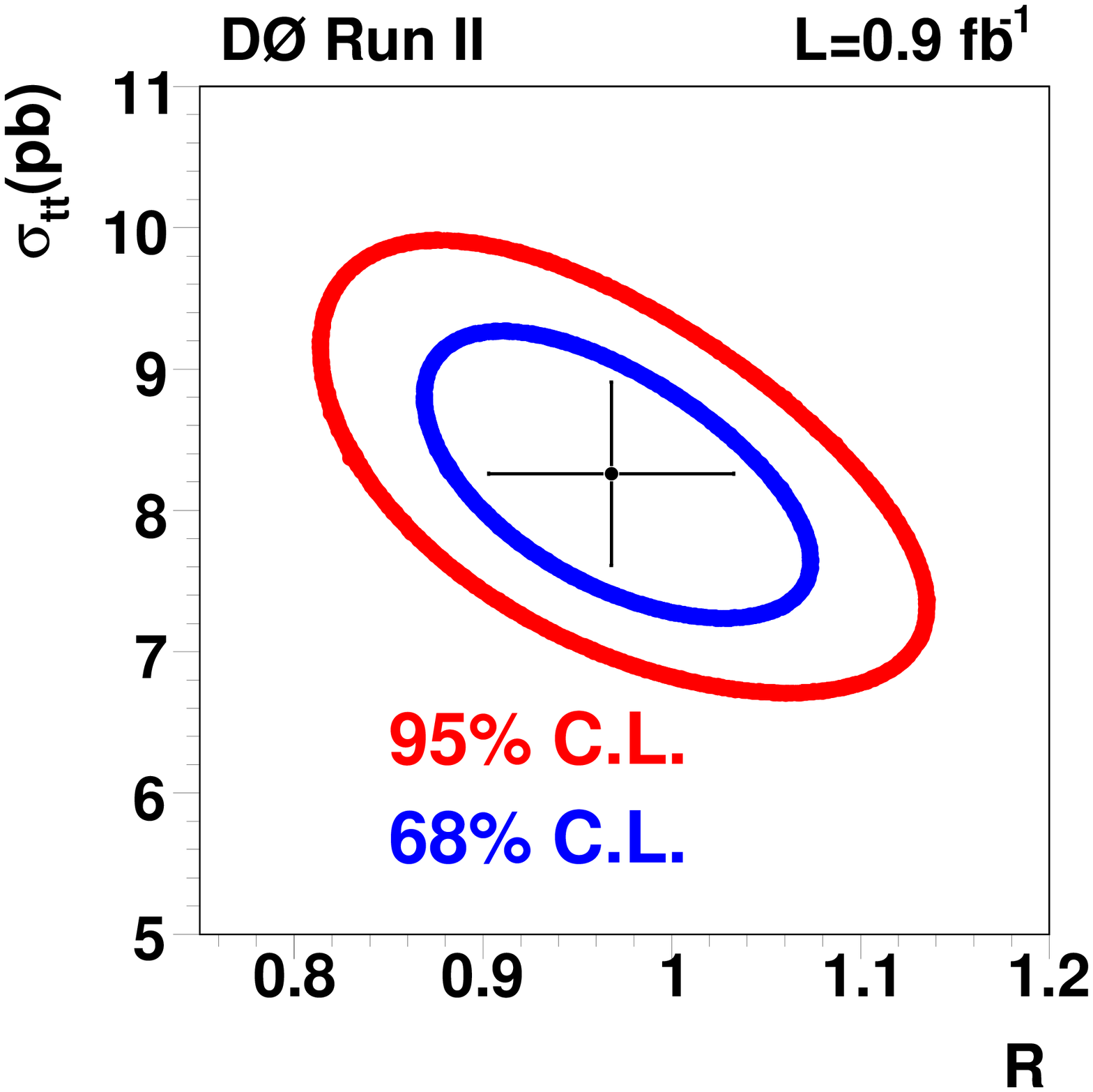} 
    \caption{\it Left: Fractions of events with 0, 1 and $\geq 2$
      $b$-tags for \ttbar events with $\geq 4$ jets as a function of
      $R$. Right: 68\% and 95\% C.L. statistical uncertainty contours
      in the R vs. $\sigma_{t\bar{t}}$ plane\cite{D0Rmeas}.}
      \label{fig:D0Rmeas}
  \end{center}
\end{figure}
Assuming the validity of the SM, specifically the existence
of three fermion generations, unitarity of
the CKM matrix and insignificance of non-$W$ boson decays of the top quark
(see Section~\ref{sec:FCNCdecay}), the ratio of branching fractions
$R$ = ${\cal{B}}(t \rightarrow Wb) / \Sigma_{q=d,s,b} {\cal{B}}(t
\rightarrow Wq$) simplifies to $|V_{tb}|^{2}$, and hence is strongly
constrained: $0.9980 < R < 0.9984$ at 90\% C.L.\cite{PDG}. Deviations
of $R$ from unity could for example be caused by the existence of a
fourth heavy quark generation.

The most precise measurement of $R$ thus far has been performed by
\dzero in the lepton+jets channel using data corresponding to an
integrated luminosity of 900~pb$^{-1}$. By comparing the event yields
with 0, 1 and 2 or more $b$-tagged jets and using a topological
discriminant to separate the \ttbar signal from background in events
with 0 $b$-tags, $R$ can be extracted together with the \ttbar
production cross section $\sigma_{t\bar{t}}$ simultaneously (see
Figure~\ref{fig:D0Rmeas}). This measurement allows the extraction of
$\sigma_{t\bar{t}}$ without assuming ${\cal B}(t\to Wb)= 100\%$,
yielding $R =0.97^{+0.09}_{-0.08}$~(stat+syst) and $\sigma_{t\bar{t}}
= 8.18^{+0.90}_{-0.84}$ (stat+syst) $\pm$ 0.50~(lumi)~pb for a top
quark mass of $175$~GeV/c$^{2}$ in agreement with the SM
prediction\cite{D0Rmeas}.

\section{Search for Flavour Changing Neutral Currents in Top Decays}
\label{sec:FCNCdecay}
\begin{figure}[!t]
  \begin{center} 
    \includegraphics[width=\textwidth, height = 50mm]{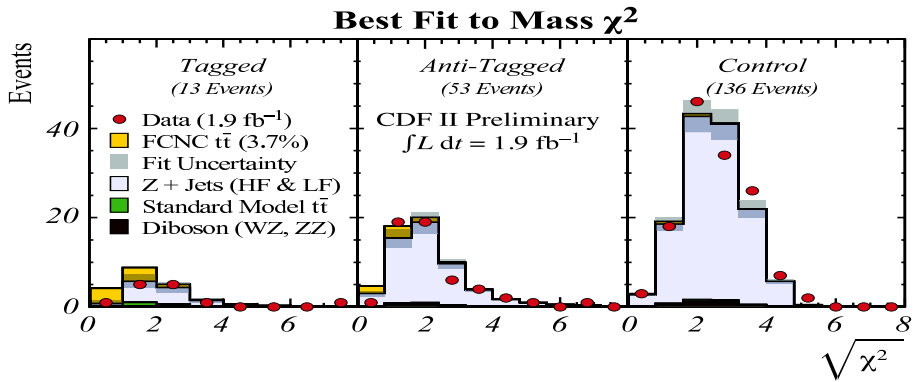}
    \caption{\it Expected and observed mass $\chi^{2}$ distributions of
      $Z + \geq 4$ jets events in signal samples with $\geq$1 and 0 $b$-tags
      and a background enriched sample to control uncertainties of the
      background shape and normalisation\cite{CDFFCNC}.}
      \label{fig:CDFFCNC}
  \end{center}
\end{figure}
The occurrence of flavour changing neutral currents (FCNC) -- a decay
of type $t\to Vq$ with $V = Z,\gamma, g$ and $q = u, c$ -- is strongly
suppressed in the SM and expected to occur at a rate below ${\cal
  O}(10^{-10})$, well out of reach of being observed at the Tevatron.
Consequently, any observation of FCNC decays would signal physics
beyond the SM.

CDF has performed a search for $t\to Zq$ in a 1.9~fb$^{-1}$ dataset of
$Z + \geq 4$ jets events with $Z\to e^{+}e^{-} \rm{or}\
\mu^{+}\mu^{-}$, assuming a SM decay of the second top quark $t\to
\overline{q}q'b$. Since the event signature does not contain any
neutrinos, the events can be fully reconstructed. The best
discriminant found to separate signal from background processes is a
mass $\chi^{2}$ variable that combines the kinematic constraints
present in FCNC decays. The signal fraction in the selected dataset is
determined via a template fit in signal samples with 0 or $\geq$1 $b$-tags
and a background-enriched control sample to constrain uncertainties on
the background shape and normalisation (see Figure~\ref{fig:CDFFCNC}).

Since the observed distributions are consistent with the SM background
processes, a 95\% C.L. upper limit on the branching fraction ${\cal
  B}(t\to Zq)$ of 3.7\% is derived\cite{CDFFCNC}. This is the best 
limit on ${\cal B}(t\to Zq)$ to date.

\section{Measurement of the $\bf{W}$ Boson Helicity in Top Quark Decays}
\label{sec:Whel}
\begin{figure}[!t]
  \begin{center} 
    \includegraphics[width=.48\textwidth, height=50mm]{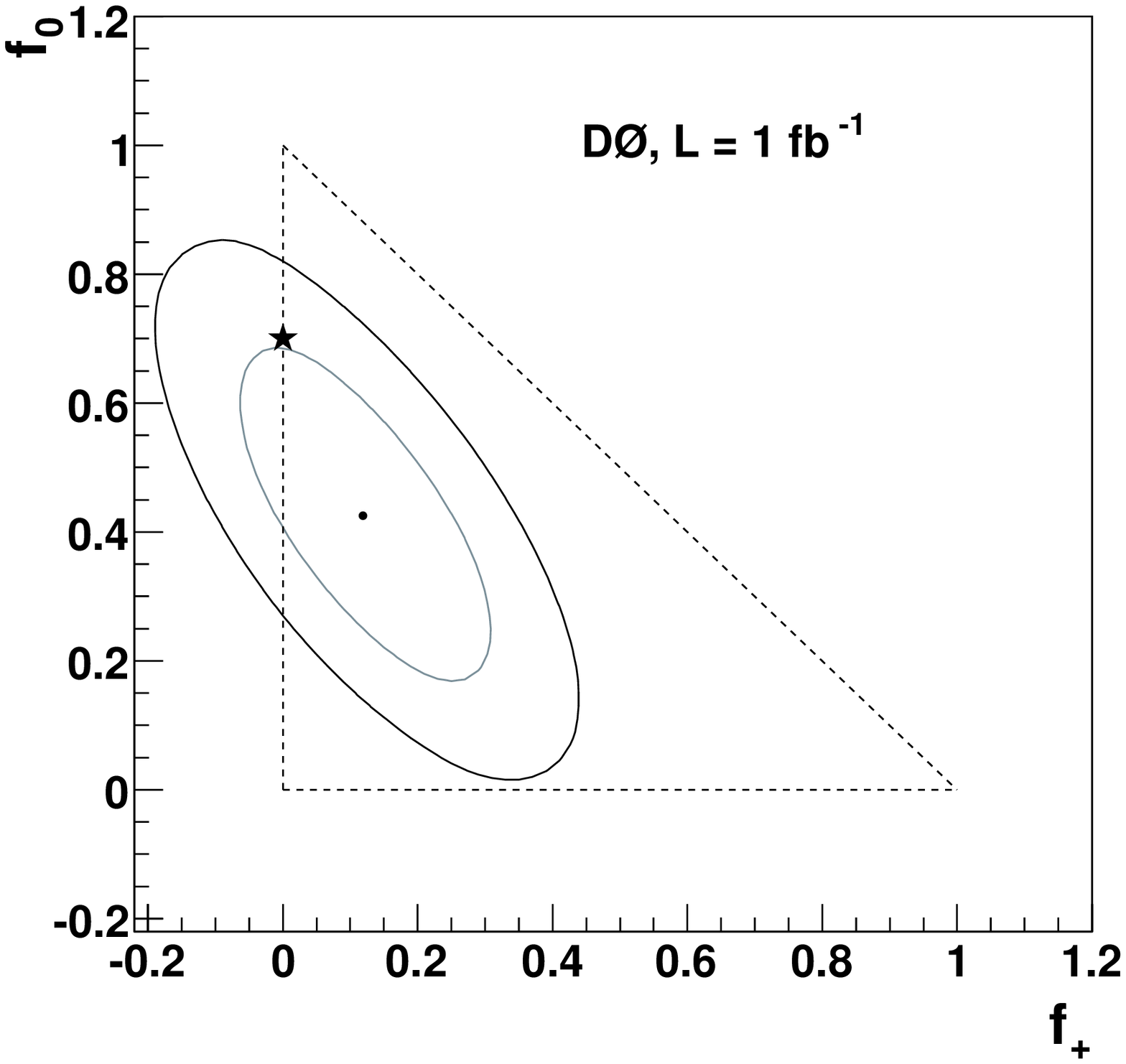}\hspace*{1mm}
    \includegraphics[width=.48\textwidth, height=50mm]{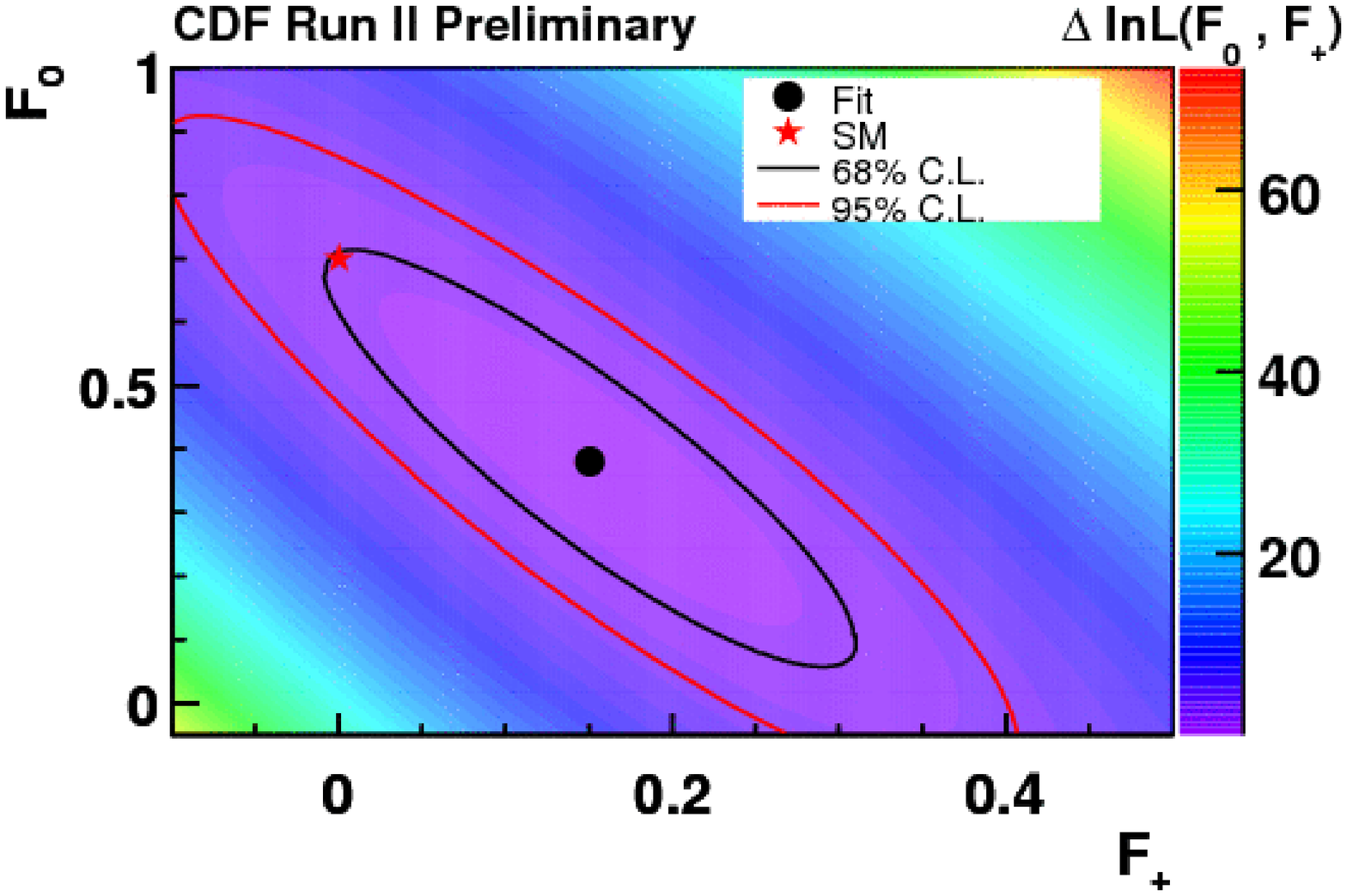} 
    \caption{\it Results of the $W$ boson helicity
      fits (left: D\O\cite{D0Whel}, right: CDF\cite{CDFWhel}). The
      ellipses show the 68\% and 95\% C.L. contours around the
      measured data points. The stars show the SM expectation; the triangle denotes the physically allowed region where $f_0$ and $f_+$ sum to one or less.}
    \label{fig:Whelmeas}
  \end{center}
\end{figure}
Assuming a massless $b$-quark, the top quark decay in the V$-$A
charged current weak interaction proceeds only via a left-handed
($f^{-}\approx$~30\%) and a longitudinal ($f^{0}\approx$~70\%) fraction
of $W$ boson helicities. The helicity of the $W$ boson is reflected in
the angular distribution $\cos\theta^{\ast}$ of its decay products,
with $\theta^{\ast}$ being the angle of the down-type decay products
of the $W$ boson (charged lepton respectively $d$- or $s$-quark) in
the $W$ boson rest frame with respect to the top quark direction. Any
observed right-handed fraction $f^{+} > \mathcal{O}(10^{-3})$ would
indicate physics beyond the SM.

CDF has measured the $W$ boson helicity fractions in 1.9~fb$^{-1}$ of
$b$-tagged lepton+jets data comparing the $\cos\theta^{\ast}$
distribution of leptons in data to templates for longitudinal, right-
and left-handed signal plus background templates. When fitting both
$f^{0}$ and $f^{+}$ simultaneously, the result is $f^{0} = 0.38 \pm
0.21\:{\rm(stat)} \pm 0.07\:{\rm(syst)}$ and $f^{+} = 0.15 \pm
0.10\:{\rm(stat)} \pm 0.05 \:{\rm(syst)}$\cite{CDFWhel}.

\dzero has measured the $W$ boson helicity fractions using the
$\cos\theta^{\ast}$ distributions in dilepton and lepton+jets events
including their hadronic $W$ boson decays in 1~fb$^{-1}$ of data,
yielding $f_0 = 0.425 \pm 0.166 \hbox{ (stat.)} \pm 0.102 \hbox{
  (syst.)}$ and $f_+ = 0.119 \pm 0.090 \hbox{ (stat.)} \pm 0.053
\hbox{ (syst.)}$\cite{D0Whel}.

Both measurements agree with the SM at the 1$\sigma$ level
(see Figure~\ref{fig:Whelmeas}).

\section{Conclusion}
\label{sec:conclusion}
A wealth of top quark analyses is being pursued at the Tevatron,
probing the validity of the SM with unprecedented precision. The
measured top quark pair production rates are found to be consistent
with the SM expectation across the decay channels, with the most
precise measurements surpassing the precision of theory predictions.
There is no evidence thus far for contributions beyond the SM in either
top quark production or top quark decay. However, with some
measurements still being statistically limited, there is still room
for surprises. More detailed descriptions of the analyses presented
here and many more interesting top quark physics results can be found
online\cite{top_group_web}.

Continuously improving analysis methods and using the increasing
integrated luminosity from a smoothly running Tevatron that is
expected to deliver more than 6~fb$^{-1}$ by the end of Run~II, we are
moving towards more precision measurements and hopefully discoveries within
and outside the SM.

\section{Acknowledgements}
The author would like to thank the organisers for creating a very
fruitful collaborative atmosphere at the Rencontres de Physique de la
Vall\'ee d'Aoste, the CDF and \dzero collaborations, the staffs at
Fermilab and collaborating institutions and also the Alexander von
Humboldt Foundation for their support.

\end{document}